\documentclass[aps,prb,floatfix,twocolumn,reprint,superscriptaddress]{revtex4}
\usepackage{amsmath}
 \usepackage{amssymb}
 
\usepackage{graphicx}
\usepackage{dcolumn}
\usepackage{bm}

\usepackage{amsfonts}
\usepackage{mathrsfs}
\usepackage{color}
\usepackage{xspace}
\usepackage{epstopdf}
\usepackage{longtable}
\usepackage{verbatim}
\usepackage{url}
\usepackage{float}

\begin{document}
\preprint{APS/123-QED}
\title{Canted Antiferromagnetism in the Quasi-1D Iron Chalcogenide BaFe$_{\bm 2}$Se$_{\bm 4}$
}
\author{Xiaoyuan Liu}
\affiliation{Department of Physics, University of Texas at Dallas, Richardson, Texas 75080, USA}
\author{Keith M. Taddei}
\affiliation{Neutron Scattering Division, Oak Ridge National Laboratory, Oak Ridge, Tennessee 37831, USA}
\author{Sheng Li}
\affiliation{Department of Physics, University of Texas at Dallas, Richardson, Texas 75080, USA}
\author{Wenhao Liu}
\affiliation{Department of Physics, University of Texas at Dallas, Richardson, Texas 75080, USA}
\author{Nikhil Dhale}
\affiliation{Department of Physics, University of Texas at Dallas, Richardson, Texas 75080, USA}
\author{Rashad Kadado}
\affiliation{Department of Physics, University of Texas at Dallas, Richardson, Texas 75080, USA}
\author{Diana Berman}
\affiliation{Department of Materials Science and Engineering, University of North Texas, Denton, Texas 76203, USA}
\author{Clarina Dela Cruz}
\affiliation{Neutron Scattering Division, Oak Ridge National Laboratory, Oak Ridge, Tennessee 37831, USA}
\author{Bing Lv}\email{blv@utdallas.edu}
\affiliation{Department of Physics, University of Texas at Dallas, Richardson, Texas 75080, USA}
\affiliation{Department of Materials Science and Engineering, University of Texas at Dallas, Richardson, Texas 75080, USA}


\begin{abstract}
We report the synthesis and physical properties studies of quais-1D iron chalcogenide $\rm BaFe_2Se_4$ which shares the $\rm FeSe_4$ tetrahedra building motif commonly seen in the iron chalcogenide superconductors. A high-quality polycrystalline sample was achieved by solid-state reaction method and characterized by X-ray diffraction, electrical resistivity, magnetic susceptibility and neutron diffraction measurements. $\rm BaFe_2Se_4$ is a narrow gap semiconductor that magnetically orders at $\sim$ 310 K. Both neutron powder diffraction results and isothermal \textit{M}-\textit{H} loops suggest a canted antiferromagnetic structure where Fe sublattice are antiferromagnetically ordered along the \textit{c}-axis quasi-1D chain direction, resulting in a net ferromagnetic moment in the perpendicular direction along the \textit{a}-axis with tilted angle of 18.7$^\circ$ towards the \textit{b}-axis.
\end{abstract}
\maketitle

The discovery of iron-based superconductors~\cite{Kamihara-06,kamihara-08} has significantly changed the landscape of unconventional superconductivity in the past decade.  Among all the iron-based superconductors, two major chalcogenide-based families have been intensively studied. First, the FeSe, with the simplest crystal structure and bulk superconductivity at 8 K at ambient pressure~\cite{Hsu-08,McQueen-09} and at 37 K under high pressure~\cite{Mizuguchi-08,Margadonna-09,Medvedev-09,Braithwaite-09,Garbarino-09}, has caught particular research interest in the past few years due to the significantly enhanced interfacial  superconductivity with $T_{\rm c}$ up to 65 K found in the epitaxial single-layer $\rm FeSe/SrTiO_3$ system~\cite{Wang-12,Liu-12,He-13,Tan-13,Lee-14,Zhang-15}. The second major family is alkali metal (A = K, Rb, Cs, and Tl) intercalated $\rm A_{1-x}Fe_{2-y}Se_2$ superconductors with $T_{\rm c}$ $\sim$ 30 K~\cite{Guo-10,Wang-11,Mizuguchi-11,Krzton-Maziopa-11,Fang-11,Qian-11,Taddei-15} where the interplay of vacancy order, magnetism, orbital-selective Mott phase, and superconductivity have been intensively studied~\cite{Taddei-15,Yi-13,Yi-15,Ye-11,Dagotto-13}. New chemical intercalation routes~\cite{Burrard-Lucas-13,Lu-15,Dong-15,Dong-2015,Zhao-16,Liu-18,Chen-19} also results in several new superconductors with higher $T_{\rm c}$ such as the $\rm Li_x(NH_2)_y(NH_3)_{1-y}Fe_2Se_2$ (x $\sim$ 0.6; y $\sim$ 0.2), and (Li,Fe)OHFeSe. Nevertheless, the essential charge carrier layers of these chalcogenide superconductors, two-dimensional (2D) $\rm Fe_2Se_2$ square lattice formed by the edge-sharing $\rm FeSe_4$ tetrahedra, are the fundamental building block of all the Fe-based superconductors. 

The spin-ladder compound $\rm BaFe_2Se_3$ is structurally related to the iron chalcogenide superconductors mentioned above, but with a reduced dimensionality. The structure consists of the same building motif, edge-sharing $\rm FeSe_4$ tetrahedra, but stacked along the \textit{b}-axis thus forming unique quasi-one-dimensional (quasi-1D) double chains of $\rm [Fe_2Se_3]$ instead of the 2D $\rm [Fe_2Se_2]$ lattice seen in the iron chalcogenide superconductors. It exhibits unique spin-ladder magnetic structure and long-range-ordered antiferromagnetic order below $T_{\rm N}$ $\sim$ 250 K, and short-range magnetic correlations at room temperature~\cite{Caron-11,Krzton-Maziopa-2011}. The antiferromagnetism changes from block-type in $\rm BaFe_2Se_3$ to stripe-type in $\rm BaFe_2S_3$, $\rm KFe_2Se_3$, and $\rm CsFe_2Se_3$~\cite{Chi-16,Caron-12,Du-12}. Interestingly, superconductivity has been reported in both $\rm BaFe_2Se_3$ and $\rm BaFe_2S_3$ compounds under high pressure, with $T_{\rm c}$ $\sim$ 11 K~\cite{Ying-17} above 10 GPa for $\rm BaFe_2Se_3$, and $T_{\rm c}$ $\sim$ 24 K above 10 GPa for $\rm BaFe_2S_3$~\cite{Yamauchi-15}, respectively. They are the only reported materials so far to exhibit a superconducting phase under pressure in this spin-ladder family. This is rather intriguing as the superconductivity in $\rm BaFe_2Se_3$ emerges near a possible orbital-selective Mott-insulator~\cite{Ying-17}, which might provide additional insight into the understanding of the 2D iron chalcogenide superconductors.  

\begin{figure}[!htbp] \centering 
\includegraphics[width=0.9\columnwidth]{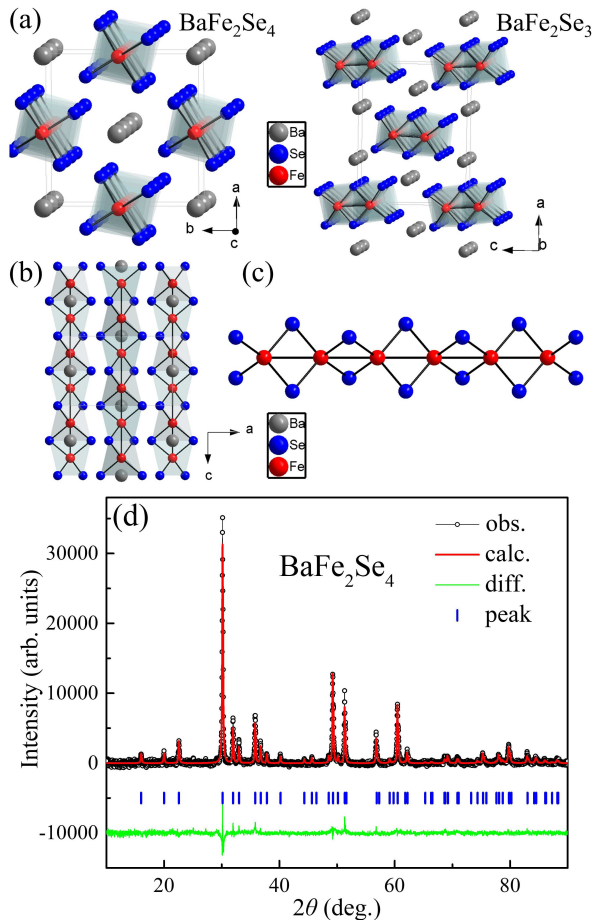}
\caption{(a) Schematic crystal structure of $\rm BaFe_2Se_4$ and $\rm BaFe_2Se_3$. Both compounds adopt similar quasi-1D structure consist of edge-sharing $\rm FeSe_4$ tetrahedra, forming single chains in $\rm BaFe_2Se_4$ while double chains in $\rm BaFe_2Se_3$ (b) the view from the \textit{b}-axis for $\rm BaFe_2Se_4$ (c) The detailed microstructure of infinite chains of edge-sharing $\rm FeSe_4$ tetrahedra along the \textit{c}-axis for $\rm BaFe_2Se_4$. (d) X-ray diffraction pattern and Rietveld refinement results of the $\rm BaFe_2Se_4$ at room temperature. The black circles are experimental data, the red curve is the Rietveld refinement fit, and the green curve is the difference. The tick marks indicate the allowed Bragg reflections.
\label{fig1}}
\end{figure}

$\rm BaFe_2Se_4$ is another new quasi-1D iron chalcogenide based on the $\rm FeSe_4$ tetrahedra building motif but with a simpler and different structure from that of the $\rm BaFe_2Se_3$ compound. The difference between the two crystal structures of $\rm BaFe_2Se_4$ and $\rm BaFe_2Se_3$ is shown in Figure~\ref{fig1}a. Unlike the spin-ladder double chains in the $\rm BaFe_2Se_3$, the quasi-1D chain in the $\rm BaFe_2Se_4$ consists only of single chains of edge-sharing $\rm FeSe_4$ tetrahedra along the \textit{c}-axis separated by Ba atoms. Figure~\ref{fig1}b and Figure~\ref{fig1}c are highlighting the quasi-1D chain along different directions. The interlayer chains are well separated from each other, with a Fe-Fe distance between the interlayer neighboring 1D Fe chains of 5.663(1) {\AA} and the closest Se-Se distance of 3.489(8) {\AA}. On the other hand, within the quasi-1D chain (Figure~\ref{fig1}c), the Fe-Fe interchain distance is 2.742(9) {\AA} and the Fe-Se distance within the $\rm FeSe_4$ tetrahedra is 2.349(5) {\AA}. Both are pretty comparable with the distances in the $\rm BaFe_2Se_3$ and other iron chalcogenide superconductors. 

\begin{table}[!htbp] \centering
\caption{\label{tab:table1}Refined atomic position at room temperature from Rietveld analysis for $\rm BaFe_2Se_4$ in the \textit{I4/m} space group with \textit{a} = 8.0150(1) $\rm \AA$, and \textit{c} = 5.4936(2) $\rm \AA$. $U_{\rm eq}$ is defined as one-third of the trace of the orthogonalized $U_{\rm ij}$ tensor.}
\begin{ruledtabular}
\begin{tabular}{cccccc}
 Atom&Wyck.&x&y&z&$U_{\rm eq}$$\rm (\AA^{2})$\\ \hline
 Ba&$2b$&$1/2$&$1/2$&$0$&$0.0411(6)$ \\
 Se&$8h$&$0.70360(6)$&$0.12350(6)$&$0$&$0.0195(3)$ \\
 Fe&$4d$&$1/2$&$0$&$1/4$&$0.0167(6)$ \\
\end{tabular}
\end{ruledtabular}
\end{table}

However, to date, only the structural determination based on small grains ($\sim$ 50 $\mu$m) has been reported for this compound~\cite{Berthebaud-15}. Previous synthesis attempts from pure elements and repeated regrinding/annealing using different temperature profiles up to 1100 $^\circ$C do not yield an X-ray powder pure phase~\cite{Berthebaud-15}. Therefore, no systematic transport and magnetic characterizations have been carried out on this compound. Band structure calculations, on the other hand, suggest the Fermi level of this compound lies near a local minimum in the density of states and an overall metallic behavior for $\rm BaFe_2Se_4$~\cite{Berthebaud-15}.  In this work, we present our studies on high quality and X-ray pure phase $\rm BaFe_2Se_4$ polycrystalline samples. In contrast to the previous calculation results, we show that the compound is actually semiconducting with an excitation energy of $\sim$ 142 meV. More importantly, it contains a unique magnetic transition at $\sim$ 310 K caused by canted antiferromagnetic interactions between the neighboring Fe-Fe atoms based on magnetic susceptibility and neutron diffraction studies. 

The $\rm BaFe_2Se_4$ polycrystalline sample was synthesized by the solid-state reaction method using elemental Ba pieces (Alfa Aesar, 99+\%), Fe granules (Alfa Aesar, 99.98\%) and Se shots (Alfa Aesar, 99.999\%). The reaction was carried out at 700 $^\circ$C in the sealed quartz tube for 2 days, followed by slowly cooling down to room temperature in 10 hours. In particular, two strategies have been utilized during the synthesis to ensure the high-quality powder synthesis and to avoid the formation of small yet magnetic $\rm FeSe_2$, $\rm Fe_7Se_8$, $\rm Fe_{1-x}Se$  and other Fe related impurities phases as commonly seen in the previous reports and other ternary iron chalcogenide phases: (1) an excess of $\sim$ 5\% Se was used to compensate the Se lost during the initial reaction from elements. (2) after the initial reaction, small representative sample is scanned by X-ray diffraction (XRD) for initial screening of phase purity where a small amount of FeSe ({\textless} 5\% based on XRD analysis) is detected. To further remove this impurity and potentially other amorphous impurities that could not been detected by XRD, a precise amount of $\rm BaSe_3$ corresponding to that needed to facilitate the reaction of $\rm BaSe_3$ + FeSe $\rightarrow$ $\rm BaFe_2Se_4$ is added to the bulk powder, and is reground together with the samples. The assembly is pelletized and annealed at slightly lower temperature 600 $^\circ$C for 2 weeks. To further improve homogeneity and quality of the samples, several cycles ({\textgreater} 2 times) of regrinding and longtime annealing ({\textgreater} 2 weeks) at 600 $^\circ$C are also carried out. A typical run of the sample normally takes more than 6 weeks to be finished. Such prolonged annealing time also exclude the formation of amorphous impurities as well. The obtained $\rm BaFe_2Se_4$ sample is stable in air, and we do not observe any impurity peaks based on XRD analysis and magnetic susceptibilities measurement.

The powder sample was characterized at room temperature by XRD using a Rigaku Smartlab diffractometer and the Rietveld refinements were carried out using FULLPROF and GSAS software packages~\cite{Rodriguez-Carvajal-93,Toby-13}. The XRD pattern of the synthesized $\rm BaFe_2Se_4$ sample and corresponding structural refinement results are presented in Figure~\ref{fig1}d and Table~\ref{tab:table1}. Based on the Rietveld analysis, the synthesized $\rm BaFe_2Se_4$ sample crystallizes in the tetragonal space group \textit{I4/m} (87) with \textit{a} = \textit{b} = 8.0150(1) {\AA}, and \textit{c} = 5.4936(2) {\AA}, consistent with the values reported in early literature~\cite{Berthebaud-15}. All diffraction peaks for $\rm BaFe_2Se_4$ can be well indexed and no impurities are detected within XRD resolution. Together with the good refinement values \textit{R} = 1.71{\%} and \textit{$R_{\rm wp}$} = 2.52{\%}, this suggests high quality of our synthesized $\rm BaFe_2Se_4$ powder sample. 

\begin{figure}[htbp] \centering 
\includegraphics[width=0.9\columnwidth]{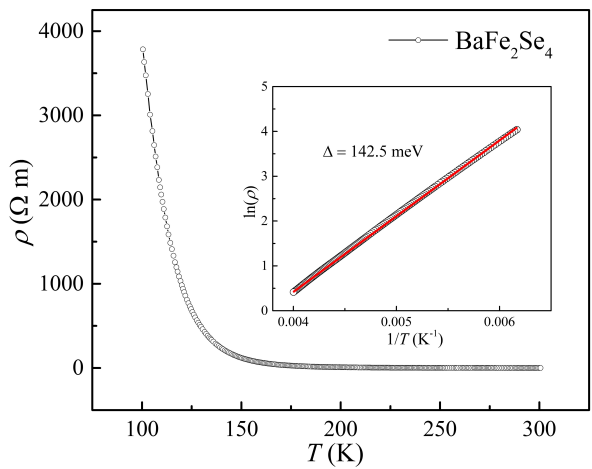}
\caption{Temperature dependence of resistivity on $\rm BaFe_2Se_4$ and the inset shows fitting results from ln(\textit{$\rho$}) vs. 1/\textit{T}.
\label{fig2}}
\end{figure}

The electrical transport was measured with Quantum Design PPMS, using the standard four-probe method. The temperature dependent electrical resistivity data are shown in Figure~\ref{fig2}. The resistivity shows an overall semiconducting behavior, which is in contradiction with previous band structure calculations~\cite{Berthebaud-15}, but in agreement with the measurement reported for its isostructural counterpart $\rm BaFe_2S_4$~\cite{Gopalakrishnan-83}. We note that the resistivity value at room temperature is 35.5 $\Omega$ cm, which is slightly larger than that of spin-ladder compound $\rm BaFe_2Se_3$ (17 $\Omega$ cm)~\cite{Lei-11}. The resistivity can be fit quite well using thermal activation model $\rho$ = $\rho_0$exp($\Delta$/$k_{\rm B}$\textit{T}), where $\rho_{\rm 0}$ is a prefactor, and $k_{\rm B}$ is the Boltzmann constant. The inset of Figure~\ref{fig2} shows the results of linear fitting of \textit{\rm ln($\rho$)} vs. 1/\textit{T}, which is consistent with the standard activation model. The activation energy estimated from the fitting is $\Delta$ = 142.5 meV, which is comparable to the reported gap value of $\rm BaFe_2Se_3$ ($\sim$ 180 meV)~\cite{Lei-11}.

\begin{figure}[htbp] \centering 
\includegraphics[width=0.9\columnwidth]{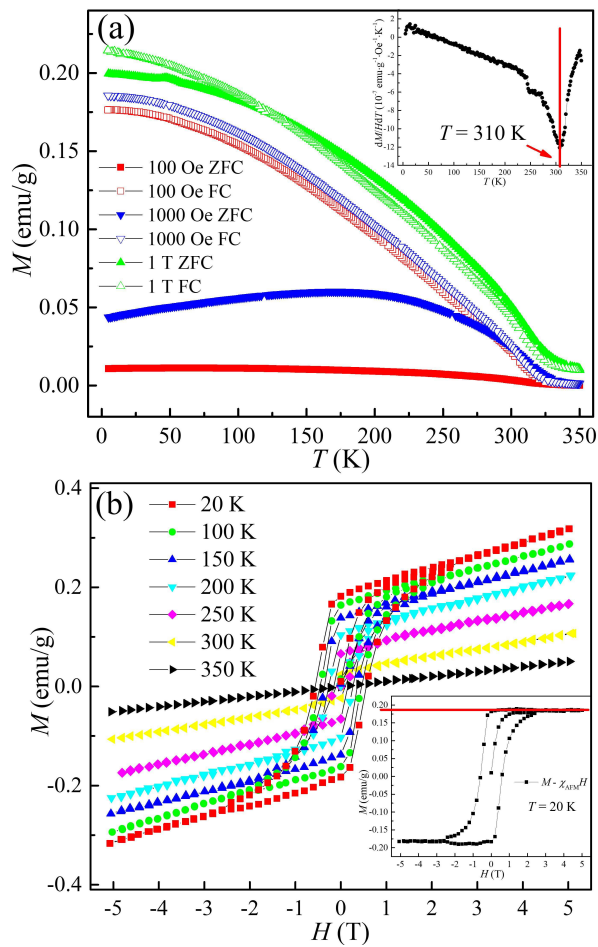}
\caption{(a) Magnetization as a function of temperature \textit{M}(\textit{T}) of $\rm BaFe_2Se_4$ measured with ZFC and FC modes at different applied magnetic fields. The inset shows the transition temperature determined from the d\textit{M}/\textit{H}d\textit{T} vs. \textit{T} plot. 
(b) Magnetization as a function of field \textit{M}(\textit{H}) of $\rm BaFe_2Se_4$ measured at different temperatures from 20 K to 350 K. The magnetic hysteresis loop is suppressed with increasing temperature and eventually disappears and becomes a straight line when temperature reaches 350 K. The inset shows the fitting data after subtracting the linear antiferromagnetic contribution \textit{$\chi_{\rm AFM}$}\textit{H} at 20 K.
\label{fig3}}
\end{figure}

The dc magnetic susceptibility and field-dependent magnetization were measured with Quantum Design MPMS down to 5 K and up to 5 T. The temperature dependence of the magnetization of $\rm BaFe_2Se_4$ from 5 K to 350 K under different magnetic fields is presented in Figure~\ref{fig3}a. A spontaneous magnetization appears at $\sim$ 310 K and increases monotonically below this transition with decreasing temperature. Together with the clear splitting of magnetization between the zero-field-cooled (ZFC) and field-cooled (FC) mode, this indicates the existence of a ferromagnetic component below the transition temperature. The $M_{\rm ZFC}$ and $M_{\rm FC}$ splitting becomes weaker with an increasing magnetic field eventually nearly overlapping at 1 T. The magnetization starts to saturate to a nearly constant plateau below about 50 K. Increasing the applied magnetic field increases the absolute value of the magnetization, but the change of transition temperature $T_{\rm c}$ is rather small ({\textless} 5 K). The exact transition temperature is tracked best in the temperature derivative of the normalized magnetization d\textit{M}/\textit{H}d\textit{T}, depicted in inset of Figure~\ref{fig3}a. No traceable magnetic transition at $\sim$ 120 K from Fe-Se binary phase ($\rm Fe_7Se_8$) is observed, further suggesting the success of our synthetic approach for high-quality samples. A small blip at $\sim$ 50 K in ZFC data at 1 T was observed which is due to some trapped oxygen in the powder samples, which is non-intrinsic signal of $\rm BaFe_2Se_4$ sample. To obtain the effective magnetic moment of Fe in $\rm BaFe_2Se_4$, we performed Curie-Weiss fitting over the high-temperature range. To eliminate the influence of ferromagnetic behavior, the temperature range between 330 K and 350 K was chosen at where the \textit{M}(\textit{T})/\textit{H} curves of 100 Oe, 1000 Oe, and 1 T overlap with each other. The susceptibility does follow the Curie-Weiss law \textit{$\chi$} = \textit{C}/(\textit{T} - $\theta$) quite nicely, with Curie constant \textit{C} = 2.82 emu K/mol and Curie temperature $\theta$ = 329 K. The resulting Curie constant corresponds to an effective magnetic moment of $\mu_{\rm eff}$ = 3.36 $\mu_{\rm B}$/Fe (4.75 $\mu_{\rm B}$/f.u.). The positive $\theta$, which is close to the transition temperature, on the other hand, indicates ferromagnetic interactions in the samples. 

The field dependence measurements of the magnetization for $\rm BaFe_2Se_4$ from -5 T to 5 T at different temperatures from 20 K to 350 K were carried out to further explore the nature of magnetic orders and are shown in Figure~\ref{fig3}b. A weak yet clear magnetic hysteresis has been observed below 310 K, however the magnetization did not saturate at a high magnetic field, instead, linear dependent \textit{M-H} curves suggest an antiferromagnetic state is observed. The magnetic hysteresis loop is suppressed with increasing temperature and eventually disappears and becomes a straight line by 350 K. No hysteresis nor clear remnant signal is observed in the isothermal \textit{M-H} loop at 350 K, suggesting the ferromagnetic components are not due to amorphous iron or iron oxide impurities, which all have a Curie temperature above 450 K. This further suggests that the observed concurrence of the ferromagnetic and antiferromagnetic state is an intrinsic property of our samples. For a high-quality pure phase, this \textit{M-H} behavior is consistent with the canted antiferromagnetic state in the $\rm BaFe_2Se_4$ system in which the ferromagnetic order is provided by the small component canted by magnetic moment of Fe.

The hysteresis loop at 20 K closes up at the field value \textit{H} $\sim$ 2.6 T and shows a coercive field of $H_{\rm c}$ = 5.6 kOe. The linear antiferromagnetic behavior appears hold up to 5 T, the highest field we have measured. To extract the saturation moment caused by the ferromagnetic component, the \textit{M}(\textit{H}) curve at the high field was fit using \textit{M}(\textit{H}) = $M_{\rm S}$ + $M_{\rm AFM}$, where \textit{$M_{\rm S}$} is the saturation moment which is field independent at the high magnetic field and \textit{$M_{\rm AFM}$} is the antiferromagnetic contribution which has a linear relationship to the magnetic field. Through this fitting, we are able to subtract the linear contribution $\chi_{\rm AFM}$\textit{H} from the experimental data, as shown in the inset of Figure~\ref{fig3}b. The moment of the ferromagnetic component could also be extracted from the saturated magnetization, from which we obtained $\sim$ 0.013 $\mu_{\rm B}$/Fe. This value is way smaller than the effective moment calculated by Curie-Weiss law, which further supports the canted antiferromagnetism scenario in $\rm BaFe_2Se_4$.

\begin{figure}[b!]
\includegraphics[width=0.9\columnwidth]{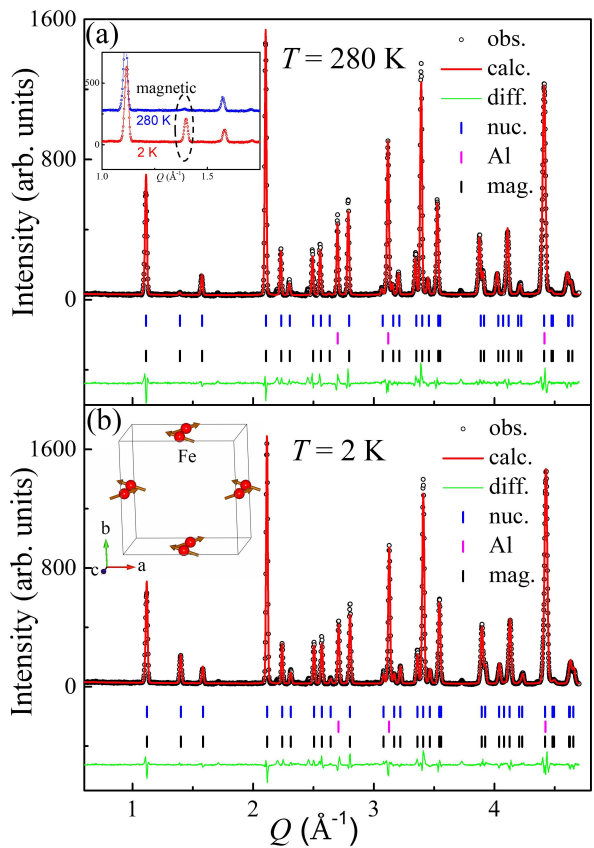}
\caption{(a) Rietveld refinement of neutron powder diffraction data of $\rm BaFe_2Se_4$ at 280 K. The inset shows the increased intensity of magnetic peak when the temperature is decreased to 2 K. (b) Rietveld refinement of neutron powder diffraction data of $\rm BaFe_2Se_4$ at 2 K. The black circles are experimental data and the red curve is the Rietveld refinement fit. The difference is shown at the bottom represented by the green curve. The positions of the nuclear and magnetic peaks of $\rm BaFe_2Se_4$ and the peaks for the aluminum sample holder are marked by the blue, black, and magenta ticks, respectively. The inset shows the schematics of the magnetic structure of $\rm BaFe_2Se_4$ (Red balls represent Fe atoms). The Fe spins are antiferromagnetically coupled along the chain direction (the $\textit{c}$-axis) and the interchain direction (the $\textit{a}$-axis), with spin along the $\textit{a}$-axis, and a small ferromagnetic canting along the $\textit{b}$-axis.
\label{fig4}}
\end{figure}

In order to gain further insights into the magnetic order, we performed temperature-dependent neutron powder diffraction (NPD). The NPD data were collected on powder sample with weight of $\sim$ 3 g using HB-2A Neutron powder diffractometer (NPDF) at Oak Ridge National Laboratory (ORNL), with Ge(113) monochromators giving wavelength 2.40 \AA~\cite{Calder-18}. The highest temperature due to experimental setup is 280 K, which below the \textit{$T_{\rm N}$}. The magnetic peak is determined through temperature dependent NPD data. As seen in the Figure~\ref{fig4}a inset, when the temperature is decreased from 280 K down to 2 K, we observe an significant increase of the magnetic Bragg intensities on top of the nuclear Bragg peaks. This suggests that the magnetic structure is resultant of a \textit{k} = (0,0,0) propagation vector which preserve the nuclear lattice’s translational symmetry. Figure~\ref{fig4}a shows a diffraction pattern taken at 280 K along with Rietveld refinement results obtained using FULLPROF~\cite{Rodriguez-Carvajal-93}. All Bragg peaks can be fit with the Rietveld refinement. The obtained refinement values $\chi^{2}$ = 6.10, \textit{$R_{\rm wp}$} = 13.1, and the \textit{R} Bragg factor for the refinement of nuclear and magnetic phases are 4.278 and 17.55, respectively. There are three additional Bragg peaks at \textit{Q} $\sim$ 2.7, 3.1, and 4.4 ${\rm \AA}^{-1}$, which belong to the aluminum sample holder and were added to the structural refinement as the minor phase. As the temperature is further decreased to 2 K, there is an increased intensity, centered at \textit{Q} $\sim$ 1.4 ${\rm \AA}^{-1}$, that is not described by the nuclear phase and likely originated from the magnetic phase.

To identify the magnetic structure, we further performed a full representational analysis for NPD data at 2 K to determine possible irreducible representations (irreps) and basis vectors (BVs) to describe the magnetic structure using the SARAh code~\cite{Wills-00}. There is a total of 6 possible basis vectors corresponding to 6 irreps \textit{$\Gamma_{\rm 1}$}, \textit{$\Gamma_{\rm 3}$}, \textit{$\Gamma_{\rm 4}$}, \textit{$\Gamma_{\rm 6}$}, \textit{$\Gamma_{\rm 7}$}, \textit{$\Gamma_{\rm 8}$}. They are labeled following the scheme of SARAh and Kovalev~\cite{Wills-00} in Table SI in the Supplemental Material~\cite{Supplemental}. Each irrep describes the magnetically distinct atoms within the unit cell. Within each irrep, each BV describes the possible direction of magnetic moments pointing along the \textit{a}, \textit{b}, and \textit{c}-axis. We discriminated between all BVs by comparing the refinement values $\chi^{2}$ and \textit{R} factors and found that both \textit{$\Psi_{\rm 3}$} and \textit{$\Psi_{\rm 5}$} are required in order to obtain the best fit, as shown in Figure~\ref{fig4}b. The \textit{$\Psi_{\rm 3}$} and \textit{$\Psi_{\rm 5}$} are from two different irreps implies that either there are two transitions or a strong first order transition. From the \textit{M-T} plots presented in Figure~\ref{fig3}a, only one transition was observed, therefore a first order transition seems more likely to be presented in $\rm BaFe_2Se_4$. The \textit{$\Psi_{\rm 5}$} indicates a ferromagnetic moment along the \textit{b}-axis, while \textit{$\Psi_{\rm 3}$} suggests an antiferromagnetic moment along the \textit{a}-axis, which further confirms the canted antiferromagnetic spin alignment in $\rm BaFe_2Se_4$. The obtained refinement values $\chi^{2}$ = 6.82 and \textit{$R_{\rm wp}$} = 12.6. The \textit{R} Bragg factors for the refinement of nuclear and magnetic phases are 4.898 and 5.758, respectively. 

The obtained magnetic structure of $\rm BaFe_2Se_4$ can be described as Fe spins aligned perpendicular to the 1D molecular chains, antiferromagnetically correlated along the chain direction (the \textit{c}-axis) as well as the interchain direction (the \textit{a}-axis), with spins along the \textit{a}-axis, and a small ferromagnetic canting along the \textit{b}-axis, as shown in the inset of Figure~\ref{fig4}b. The magnitude of the magnetic moment estimated from NPD data is 2.09 $\mu_{\rm B}$/Fe, with 1.98 $\mu_{\rm B}$/Fe along the \textit{a}-axis and 0.67 $\mu_{\rm B}$/Fe along the \textit{b}-axis. The magnetic moment canting angle is $\sim$ 18.7$^\circ$ from the \textit{a}-axis. Because of thermal fluctuations, a larger tilted angle of 63.1$^\circ$ is observed at 280 K comparing to 18.7$^\circ$ at 2 K. For comparison, in the tetrahedrally coordinated system, due to the crystal field effect, the magnetic moment of a free $\rm Fe^{3+}$ ion is 5.92 $\mu_{\rm B}$ for the high spin state and 1.73 $\mu_{\rm B}$ for the low spin state taking the Lande factor \textit{g} = 2. The calculated magnetic moment is in between the high spin and low spin state indicating a possible mixed state in this system. The magnitude of the magnetic moment is pretty close to that of single-chain quasi-1D $\rm TlFeSe_2$ material with the same formal $\rm Fe^{3+}$ valence, but smaller than that of spin-ladder quasi-1D iron chalcogenide $\rm BaFe_2Se_3$ with a formal $\rm Fe^{2+}$ valence. In $\rm TlFeSe_2$, the $\rm FeSe_4$ tetrahedra also form a quasi-1D structure with N\'{e}el temperature at 295 K and the magnetic moment of Fe is 2.1 $\mu_{\rm B}$/Fe~\cite{Seidov-01,Asgerov-15,Ismayilova-17,Seidov-17,Asgerov-18}. The similar single 1D chain in both $\rm BaFe_2Se_4$ and $\rm TlFeSe_2$, leads one to expect that both materials have similar direct and indirect exchange interactions. However, the Dzyaloshinskii-Moriya interaction also arises, and this anisotropic exchange interaction leads to the canted antiferromagnetic ground state in the $\rm BaFe_2Se_4$. On the other hand, this canted antiferromagnetic structure in $\rm BaFe_2Se_4$ is quite different from the typical magnetic structures discovered in spin-ladder quasi-1D iron chalcogenides, such as block-type structure in $\rm BaFe_2Se_3$, where four Fe spins along the chain form a $\rm Fe_{4}$ ferromagnetic block, and each $\rm Fe_{4}$ block stacks antiferromagnetically~\cite{Caron-11,Krzton-Maziopa-2011}, or stripe-type structure in $\rm BaFe_2S_3$, $\rm KFe_2Se_3$ and $\rm CsFe_2Se_3$, where Fe spins are ferromagnetic coupled along the rung direction and each ferromagnetic unit stacks antiferromagnetically along the leg direction~\cite{Chi-16,Caron-12,Du-12}.

The observation of the canted antiferromagnetism in this quasi-1D $\rm BaFe_2Se_4$ compound is rather intriguing and could be a playground to further explore the correlations between magnetism and superconductivity. The antiferromagnetic order and/or magnetic spin fluctuations have been universally observed in the iron-based superconductors, and plays an important role for the emergence of superconductivity~\cite{Allred-16,Basov-11,Fernandes-16}. $\rm BaFe_2Se_4$, with a simpler structure and higher symmetry than spin-ladder phase $\rm BaFe_2Se_3$, will provide a unique opportunity to reveal the intimate interplay between magnetism, crystal lattice and electronic structure in [$\rm FeX_4$]-based materials, and perhaps to understand the mechanism of superconductivity in Fe-based compounds. The $\rm BaFe_2Se_3$ becomes superconducting at $\sim$ 11 K at high pressure {\textgreater} 10 GPa. The appearance of superconductivity in $\rm BaFe_2Se_3$ has a strong correlation with the magnitude of magnetic moments of Fe atoms, and the magnitude of magnetic moments is gradually decreasing with increasing pressure~\cite{Ying-17}. It will be interesting to investigate how the magnetic structure and moment, canted antiferromagnetic correlation, and ferromagnetic component evolve with chemical doping or high pressure, whether superconductivity could be induced in the vicinity of this canted antiferromagnetism and how it interplays with the antiferromagnetism or even ferromagnetism when it emerges. 

In summary, we have successfully synthesized high-quality polycrystalline sample of quasi-1D iron chalcogenide $\rm BaFe_2Se_4$. It has a semiconducting behavior with activation energy $\sim$ 142 meV. Magnetic susceptibility measurements suggest a ferromagnetic-like transition at $\sim$ 310 K, and NPD further reveals it is actually canted antiferromagnetism correlated along the 1D chain direction with a net small ferromagnetic moment in the perpendicular direction. The magnitude of the magnetic moment estimated from NPD data is 2.09 $\mu_{\rm B}$/Fe, which is pretty close to that of quasi-1D $\rm TlFeSe_2$ material with the same $\rm Fe^{3+}$ valence, but smaller than that of other quasi-1D iron chalcogenides such as $\rm BaFe_2Se_3$ with $\rm Fe^{2+}$ valence. This canted antiferromagnetic structure is also different from the typical block-type or stripe-type magnetism discovered in the quasi-1D iron chalcogenide such as $\rm BaFe_2Se_3$ and $\rm BaFe_2S_3$, and could provide unique playground to study the interplay between magnetism, crystal lattice and electronic structure in the Fe-based compounds.

This work at University of Texas at Dallas is supported by US Air Force Office of Scientific Research (FA9550-19-1-0037), and National Science Foundation (DMR 1921581). We also acknowledge the support from the Office of Research at University of Texas at Dallas through the Seed Program for Interdisciplinary Research (SPIRe) and the Core Facility Voucher Program. Support from Advanced Materials and Manufacturing Processes Institute (AMMPI) at the University of North Texas is acknowledged.

\bibliography{references}
\end{document}



\title{Supplemental Material for `Canted Antiferromagnetism in the Quasi-1D Iron Chalcogenide BaFe$_{\bm 2}$Se$_{\bm 4}$'}
\author{Xiaoyuan Liu}
\affiliation{Department of Physics, University of Texas at Dallas, Richardson, Texas 75080, USA}
\author{Keith M. Taddei}
\affiliation{Neutron Scattering Division, Oak Ridge National Laboratory, Oak Ridge, Tennessee 37831, USA}
\author{Sheng Li}
\affiliation{Department of Physics, University of Texas at Dallas, Richardson, Texas 75080, USA}
\author{Wenhao Liu}
\affiliation{Department of Physics, University of Texas at Dallas, Richardson, Texas 75080, USA}
\author{Nikhil Dhale}
\affiliation{Department of Physics, University of Texas at Dallas, Richardson, Texas 75080, USA}
\author{Rashad Kadado}
\affiliation{Department of Physics, University of Texas at Dallas, Richardson, Texas 75080, USA}
\author{Diana Berman}
\affiliation{Department of Materials Science and Engineering, University of North Texas, Denton, Texas 76203, USA}
\author{Clarina Dela Cruz}
\affiliation{Neutron Scattering Division, Oak Ridge National Laboratory, Oak Ridge, Tennessee 37831, USA}
\author{Bing Lv}
\affiliation{Department of Physics, University of Texas at Dallas, Richardson, Texas 75080, USA}
\affiliation{Department of Materials Science and Engineering, University of Texas at Dallas, Richardson, Texas 75080, USA}

\maketitle

\makeatletter
\renewcommand{\tablename}{Table S\@gobble}
\makeatother

\begin{table*}[htbp]
\caption{\label{tab:table2}Basis vectors (BVs) of irreducible representations (irreps) for the space group \textit{I4/m} with the magnetic propagation vector \textit{k} = (0, 0, 0)}
\begin{ruledtabular}
\begin{tabular}{cccccc}
 Irreps&BVs&&\multicolumn{3}{c}{Basis vector components}\\ \hline
 $\Gamma_{\rm i}$&$\Psi_{\rm i}$&Atoms&$m_a$&$m_b$&$m_c$\\ \hline
 $\Gamma_{\rm 1}$&$\Psi_{\rm 1}$&Fe1&$0$&$0$&$4$ \\
 $$&$$&Fe2&$0$&$0$&$4$ \\
 $\Gamma_{\rm 3}$&$\Psi_{\rm 2}$&Fe1&$4$&$0$&$0$ \\
 $$&$$&Fe2&$4$&$0$&$0$ \\
 $\Gamma_{\rm 4}$&$\Psi_{\rm 3}$&Fe1&$4$&$0$&$0$ \\
 $$&$$&Fe2&$-4$&$0$&$0$ \\
 $\Gamma_{\rm 6}$&$\Psi_{\rm 4}$&Fe1&$0$&$0$&$4$ \\
 $$&$$&Fe2&$0$&$0$&$-4$ \\
 $\Gamma_{\rm 7}$&$\Psi_{\rm 5}$&Fe1&$0$&$-4$&$0$ \\
 $$&$$&Fe2&$0$&$-4$&$0$ \\
 $\Gamma_{\rm 8}$&$\Psi_{\rm 6}$&Fe1&$0$&$4$&$0$ \\
 $$&$$&Fe2&$0$&$-4$&$0$ \\
\end{tabular}
\end{ruledtabular}
\end{table*}
